\documentstyle[12pt]{article}
\begin{document}
\begin{center}
\vspace{1.5in}
{\LARGE
Multi - nucleon forces and the equivalence of different 
multi - cluster structures}
\end{center}
\vspace{.4in}
\begin{center}
{\bf AFSAR ABBAS}\\
Institute of Physics\\ 
Bhubaneshwar-751005, India\\
email: afsar@iopb.res.in
\end{center}
\vspace{1.2in}
\begin{center}
{\bf Abstract}
\end{center}
\vspace{.3in}

At present it is believed that different multi-cluster structures 
for light nuclei may represent equivalent states of these nuclei.
Here it is proven that once we take account of Three Body and 
Four Body Forces then this equivalence is lost.

\newpage

For quite sometime now it has been becoming clear  
that the best microscopic
calculations using successful Two Body Forces (2BF) for light nuclei A=3,4 
give underbinding and too large a radii [1]. This pointed at the 
requirement of other forces like the Three Body Forces (3BF)
and possible Four Body Forces (4BF). The improvements arising from the 
inclusion of 3BF are well known [2]. It has also been becoming clear that
4BF are essentail for a good undertanding of light nuclei [3].

Let us accept that 3BF and 4BF are needed in a basic way for the study of 
light nuclei [2,3]. It is important to note that these are genuine 3BF 
and 4BF which are irreducible to any 2BF etc. As nucleons have finite size 
and structure it is clear that there shall be genuine multi-body forces.
These will manifest themselves in any physical situation where it would be 
essential to consider overlap of multi-nucleons [4-6]. 
However, how and why 
these multi-body forces arise is not the issue here. For our purpose here 
it is enough to accept the existence of 3BF and 4BF forces. This means 
that to get good description of A=3,4 nuclei we do need 3BF and 4BF in a 
basic way to fit to experimental quantities like binding energies,
radii, charge distributions and excitation spectra of these nuclei. 

Let the Hamiltonian for upto A=4 system be
\begin{equation}
H =\sum_{i}^{A}\frac{p_{i}^{2}}{2m}+\sum_{i<j}^{A}v_{ij}+
\sum_{i<j<k}^{A}V_{ijk}+\sum_{i<j<k<l}^{A}V_{ijkl}
\end{equation}
where $v_{ij}$, $V_{ijk}$ and $V_{ijkl}$
are NN, NNN and NNNN potentials respectinely. For A=2 system only the 
first two terms would be present. While for A=3 system the NNN term would 
also be contributing. Only for A=4 system does the 4BF term 
contribute.

Now let us use this Hamiltonian to solve that A=2,3,4 body problems in
nuclear physics. We may use the Quantum Monte Carlo method or some other 
method to obtain the best fits to all the relevant experimental 
quantities of these nuclei which are: 
${^2}H$, ${^3}H$ and ${^3}He$ and ${^4}He$.
Let us assume that we are able to do as good a job 
as possible. At the end of the day let the wave functions which provided
these fits be
 $\psi_{0} (d)$,
 $\psi_{0} (t)$,
 $\psi_{0} (h)$ and
 $\psi_{0} (\alpha)$.
( Note : d=
${^2}H$, t=${^3}H$, h=${^3}He$ and $\alpha$=${^4}He$ )

Note that all these wave functions are solutions of Hamiltonian
which have componnents not present in all of these. For example the A=4 
case the wave function has " wounds " inflicted by the 4BF which are not 
present in the others. As such the wave function for A=4 case has 
components which can not be reduced to any (2-body)-(2-body)
or (3-body)-(1-body) wave functions. In the same manner the A=3 case have 
" wounded " wave functions unique to themselves 
and irreducible to any other system of wave 
functions. These 3BF and 4BF "wounds" in the wave functions in a way 
are unique signatures of their presence in A=3 and 4 systems.
Note that the A=2 case would have none of the "wounds" arising from 3BF or 
4BF.

At present it is commonly believed that there exists equivalence
of various multi-clusters for light nuclei. For example
in the oscillator model the ground state of ${^6}Li$ is described by the 
following wave functions of equivalent forms [7,p.40, appendix C] 

\begin{equation}
 \psi_{gs} (^{6}Li) = A \{ \phi_{0}(\alpha) \phi_{0}(d)
                        \chi(\alpha - d) \}
 = N A ( \phi_{0} (t) \phi_{0} (h) \chi(t-h) \}
\end{equation}

Here $\chi (\alpha-d)$ and $\chi (t-h)$ are two oscillator quanta wave
functions and
$\phi_{0} (\alpha), \phi_{0} (d), \phi_{0} (t), \phi_{0} (h) $   
are the internal wave functions. The constant N
is here to ensure that the two functions are equal to each other. The
mathematical equivalence of the two can be easily demonstrated.  Since
they are the same, these can exist simultaneously 
and also note that these are not
orthogonal [7]. None of this is changed even with the hard core of the
Jastrow form in the two body interaction [7, p.113]. This is the broad 
view which has been dominating the studies of light nuclei
at present [7-9]. 

However, in the above mathematical proof, there is an underlying ansatz 
- which is that in the Schroedinger equation one need not go beyond 
2BF and ignore all many body forces like the 
3BF and 4BF [7, p.1]. However at present on the basis of what we 
we have discussed above there is no way that one can do 
without 3BF and 4BF for these nuclei. Also due to the reasons indicated 
above the projection of the wave function

 $ \phi_{0}(\alpha) \phi_{0}(d) $

on 

$  \phi_{0} (t) \phi_{0} (h) $

would be zero.
Hence the putative equivalence of different multi-clusters [7-9]
for light nuclei and
which forms the backbone of studies of nuclei at present is wrong and 
should be abandoned.

\newpage 

\vspace{.2in}

{\bf References} 

\vspace{.3in}

1. B. F. Gibson, Nucl. Phys. {\bf A 543} (1992) 1c

\vspace{.1in}

2. B. S. Pudliner et al, Phys. Rev. {\bf C 56} (1997) 1720

\vspace{.1in}

3. F. Gross et al, Nucl. Phys. {\bf A 689} (2001) 573c

\vspace{.1in}

4. A. Abbas, Phys. Lett. {\bf B167} (1986) 150. 

\vspace{.1in}

5. A. Abbas, Prog. Part. Nucl. Phys. {\bf 20} (1988) 181.

\vspace{.1in}

6. A. Abbas, Mod. Phys. Lett. {\bf A16} (2001) 755.

\vspace{.1in}

7. K. Wildermuth and W. McClure, Springer Tracts in Modern Physics, 
{\bf 41} (1966),  Springer Verlag, Berlin. 

\vspace{.1in}

8. R.D. Amado and J.V. Noble, Phys. Rev. {\bf C3} (1971) 2494.

\vspace{.1in}

9. J.P. Connelly et al, Phys. Rev. {\bf C57} (1998) 1569.

\end{document}